\def\rfr#1{eq. (\ref{#1})}

\def\asec{$''$ cy$^{-1}$}

\def\dert#1#2{\frac{{{d}}{#1}}{{{d}}{#2}}}              

\def\asec{$''$ cy$^{-1}$}

\def\bar{\begin{eqnarray}}
\def\ear{\end{eqnarray}}
\def\bb{\bibitem}
\def\eqi{\begin{equation}}
\def\eqf{\end{equation}}
\def\eqia{\begin{eqnarray}}
\def\eqfa{\end{eqnarray}}
\def\rp#1#2{{#1\over#2}}

\def\lb#1{\label{#1}}

\def\oc2{$\mathcal{O}(c^{-2})$}

\documentclass{aastex}
\usepackage{url}

\RequirePackage{color}

\begin{document}

\title{Advances in the measurement of the Lense-Thirring effect with planetary motions in the field of the Sun}

\shorttitle{Advances in the measurement of the solar Lense-Thirring effect}
\shortauthors{L. Iorio }

\author{Lorenzo Iorio }
\affil{INFN-Sezione di Pisa. Permanent address for correspondence: Viale Unit\`{a} di Italia 68, 70125, Bari (BA), Italy. E-mail: lorenzo.iorio@libero.it}

\begin{abstract}
E.V. Pitjeva, by processing more than 400,000 planetary observations of various types with the dynamical models of the EPM2006 ephemerides, recently estimated a correction to the canonical Newtonian-Einsteinian Venus' perihelion precession of $-0.0004\pm 0.0001$ arcseconds per century. The prediction of general relativity for the Lense-Thirring precession of the perihelion of Venus is $-0.0003$  arcseconds per century. It turns out that neither other mismodelld/unmodelled standard Newtonian/Einsteinian effects nor exotic ones, postulated to, e.g., explain the Pioneer anomaly, may have caused the determined extra-precession of the Venus orbit which, thus, can be reasonably attributed to the gravitomagnetic field of the Sun, not modelled in the routines of the EPM2006 ephemerides. However, it must be noted that the quoted error is the formal, statistical one; the realistic uncertainty might be larger. Future improvements of the inner planets' ephemerides, with the inclusion of the Messenger and Venus-Express tracking data, should further improve the accuracy and the consistency of such a test of general relativity which would also benefit of the independent estimation of the extra-precessions of the perihelia (and the nodes) by other teams of astronomers.
\end{abstract}



In the weak-field and slow motion approximation the Einstein field equations of general relativity get linearized resembling  the Maxwellian equations of electromagnetism. As a consequence, a gravitomagnetic field arises \citep{Mash01,MashNOVA}; it is induced by the off-diagonal components $g_{0i}, i=1,2,3$ of the space-time metric tensor related to the mass-energy currents of the source of the gravitational field. It affects  orbiting test particles, precessing gyroscopes, moving clocks and atoms and propagating electromagnetic waves \citep{Rug,Scia04}. The most famous gravitomagnetic effects are, perhaps, the precession  of the axis of a gyroscope \citep{Pugh,Schi}, whose detection in the gravitational field of the rotating Earth is the goal of the space-based GP-B experiment\footnote{See on the WEB http://einstein.stanford.edu/} \citep{Eve}, and the Lense-Thirring\footnote{According to an interesting historical analysis recently performed in \citep{Pfi07}, it would be more correct to speak about an Einstein-Thirring-Lense effect.}  precessions \citep{LT} of the orbit of a test particle for which some disputed satellite-based tests in the gravitational fields of the spinning Earth \citep{Ciu04,Ciu05,IorNA,IorJoG,Ior07,Luc05} and Mars \citep{IorMGS,KroghMGS,IorMGSReply} have been reported.

 We focus on the detection of the solar gravitomagnetic field through the Lense-Thirring planetary precessions of the longitudes of perihelia\footnote{Here $\omega$ is the argument of pericentre, reckoned from the line of the nodes, $i$ is the inclination of the orbital plane to the equator of the central rotating mass and $\Omega$ is the longitude of the ascending node. } $\varpi=\omega+\cos i\ \Omega$
\eqi\dert\varpi t = -\rp{ 4GS\cos i }{ c^2 a^3 (1-e^2)^{3/2} },\eqf where $G$ is the Newtonian gravitational constant, $S$ is the proper angular momentum of the Sun, $c$ is the speed of light in vacuum, $a$ and $e$ are the semimajor axis and the eccentricity, respectively, of the planet's orbit.  It may be interesting to know that in \citep{Haas} it was proposed to measure the solar gravitomagnetic field through the Schiff effect with a  drag-free gyroscope orbiting the Sun in a polar orbit.

The impact of the Sun's rotation on the
Mercury's longitude of perihelion was calculated for the first
time with general relativity by \citet{deS16} who, by assuming a homogenous and
uniformly rotating Sun, found a secular rate of $-0.01$
arcseconds per century (\asec\ in the following). This value is
also quoted at pag. 111 of \citet{Sof89}. \citet{Cug78}
yield $-0.02$ \asec\ for the argument of perihelion of Mercury.
Instead, recent determinations of the Sun's proper angular
momentum\footnote{It could me measured also with a different approach \citep{Ni08}.} $S_{\odot} = (190.0\pm 1.5)\times 10^{39}$ kg m$^2$ s$^{-1}$ from helioseismology \citep{Pij98,Pij03}, accurate to $0.8\%$, yield a
precessional effect one order of magnitude smaller.
The predicted
gravitomagnetic precessions of the four inner planets, according to the recent value of the Sun's angular momentum, are reported in Table \ref{tavola1}; they are of the order of
$10^{-3}-10^{-5}$ \asec.
\begin{table*}
\small
\caption{ Lense-Thirring precessions, in \asec, of the longitudes of the perihelion $\varpi$ of the inner planets of the Solar System induced by the gravitomagnetic field of the Sun. The value $S_{\odot} = (190.0\pm 1.5)\times 10^{39}$ kg m$^2$ s$^{-1}$ has been assumed for its angular momentum.\label{tavola1}}

\begin{tabular}{@{}cccc@{}}
\hline
Mercury & Venus & Earth & Mars\\
\tableline
$-0.0020$ & $-0.0003$ & $-0.0001$ & $-0.00003$\\
\hline
\end{tabular}
\end{table*}
Due to their extreme smallness it has been believed for a long time, until recently, that the planetary Lense-Thirring effect would have been undetectable; see, e.g., p. 23 of \citet{Sof89}.
A preliminary analysis showing that recent advances in the ephemerides field are making the situation more favorable was carried out in \citep{IorAA}.
\citet{Pit05} processed more than 317,000 planetary observations of various kinds collected from 1917 to 2003 with the dynamical force models of the EPM2004 ephemerides \citep{PitSS} producing a global solution in which she estimated, among many other parameters, also corrections $\Delta\dot\varpi$ to the canonical Newton-Einstein perihelion precessions for all the inner planets; since the gravitomagnetic force was not modelled at all, contrary to the static part of the general relativistic force of order $\mathcal{O}(c^{-2})$, such corrections to the usual perihelia evolutions account, in principle, for the Lense-Thirring effect as well, in addition to the mismodelled parts of the standard Newtonian/Einsteinian precessions. Thus, the estimated corrections for the perihelion rates of Mercury, the Earth and Mars have been used in \citep{IorPS} to perform a first test. The errors $\delta(\Delta\dot\varpi)$ released in \citep{Pit05} were slightly larger than the gravitomagnetic precessions whose predicted values, however, were found compatible with the estimated corrections. Venus was not used because of the poor data set used in the estimation of its extra-precession whose value, indeed,  turned out too large to be due to a physically plausible effect amounting to $+0.53\pm 0.30 $\asec: the Lense-Thirring prediction for the Venus perihelion precession was incompatible with such a result at about $2-\sigma$ level.

Now the situation for the second planet of the Solar System has remarkably improved allowing for a more stringent test of the Lense-Thirring effect.
Indeed, \citet{Pit07a,Pit07b}, in the effort of continuously improving the planetary ephemerides, recently processed more than 400,000 data points (1913-2006) with the EPM2006 ephemerides which encompasses better dynamical models with the exception, again, of the gravitomagnetic force itself.
Also in this case she estimated, among more than 230 parameters, the corrections to the usual perihelion precessions  for some planets \citep{Pit07a}. In the case of Venus the inclusion of the radiometric data of Magellan \citep{Pit07b} as well allowed her to obtain\footnote{Personal communication by Pitjeva to the author, June 2008.}
\eqi\Delta\dot\varpi_{\rm Venus} = -0.0004\pm 0.0001\ ''\ {\rm cy}^{-1},\lb{venu}\eqf
in which the quoted uncertainty is the formal, statistical one. By looking at Table \ref{tavola1} it turns out that such an extra-precession can be well accommodated by the general relativistic prediction for the Lense-Thirring rate of the Venus'perihelion whose existence would, thus, be confirmed at $25\%$. Somebody may object that the gravitomagnetic force should have  been explicitly modelled and an ad-hoc parameter accounting for it should have been inserted in the set of parameters to be estimated. Certainly, it may be an alternative approach which could be implemented in future; in addition, we note that the procedure followed by Pitjeva  may be regarded, in a certain sense, as safer for our purposes because it is truly model-independent and, since her goal in estimating $\Delta\dot\varpi$ was not the measurement of the Lense-Thirring effect, there is a priori no risk that, knowing in advance the desired answer, something was driven just towards the expected outcome.

The main question to be asked is, at this point, the following one: Can the result of \rfr{venu} be explained by other unmodelled/mismodelled canonical or non-conventional dynamical effects?
 Let us, first, examine some standard candidates like, e.g., the residual precession due to the still imperfect knowledge of the Sun's quadrupole mass moment $J_2^{\odot}$ \citep{Roz03} whose action was, in fact, modelled by \citet{Pit05} by keeping it fixed to $J_2^{\odot}=2\times 10^{-7}$ in the global solution in which she estimated the corrections to the perihelion precessions. The answer is negative since the Newtonian secular precession due to the Sun's oblateness\footnote{For an oblate body $J_2>0$.}, whatever magnitude $J_2$ may have, is positive. Indeed, it is \citep{Capde,IorPS}
\eqi\dot\varpi_{J_2}=\rp{3}{2}\rp{nJ_2}{(1-e^2)^2}\left(\rp{R}{a}\right)^2\left(1-\rp{3}{2}\sin^2 i\right),\eqf where $n=\sqrt{GM/a^3}$ is the Keplerian mean motion and  $R$ is the Sun's mean equatorial radius; the angle $i$ between the Venus'orbit and the Sun's equator amounts to\footnote{Indeed, the orbit of Venus is tilted by 3.7 deg to the mean ecliptic of J2000 (http://ssd.jpl.nasa.gov/txt/aprx$\_$pos$\_$planets.pdf), while the Carrington's angle between the Sun's equator and the ecliptic is 7.15 deg \citep{Carr}.} 3.4 deg only. For $J_2^{\odot}=2\times 10^{-7}$, the nominal value of the Venus' perihelion precession induced by it amounts to $+0.0026$ \asec; by assuming an uncertainty of about $\delta J_2\approx 10\%$ \citep{Fien08}, if $\Delta\dot\varpi_{\rm Venus}$ was due to such a mismodelled effect it should amount to $+0.0002$ \asec, which is, instead, ruled out at $6-\sigma$ level.
Concerning the precession due to the solar octupole mass moment $J_4^{\odot}$,
it is  \citep{Capde}
\eqi\dot\varpi_{J_4}= -\rp{15}{16}n J_4\left(\rp{R}{a}\right)^4\left[\rp{3}{(1-e^2)^3} +7\rp{(1+\rp{3}{2}e^2)}{(1-e^2)^4}  \right]\left(\rp{7}{4}\sin^4 i - 2\sin^2 i +\rp{2}{5}\right).\eqf For Venus it amounts to $-1.2\ J_4^{\odot}\ \ ''\ {\rm cy}^{-1}$; since $J_4^{\odot}\approx -4\times 10^{-9}$ \citep{Rox01,Mec04}, we conclude that the second even zonal harmonic of the multipolar expansion of the solar gravitational potential cannot be responsible for \rfr{venu} and, more generally, it does not represent a potentially relevant source of systematic error for the measurement of the Lense-Thirring planetary precessions.
Similar arguments hold also for other potential sources of systematic errors like, e.g., the asteroid ring and the Kuiper Belt objects, both modelled in EPM2006: the precessions induced by them are positive. Indeed, a Sun-centered ring of mass $m_{\rm ring}$ and inner and outer radius $R_{\rm min/max}\gg a$ induces a perihelion precession \citep{IorKBO}
\eqi \dot\varpi_{\rm ring} = \rp{3}{4}\sqrt{\rp{G a^3(1-e^2)}{M}}\rp{m_{\rm ring}}{R_{\rm min}R_{\rm max}(R_{\rm min} + R_{\rm max})}>0.\lb{ring}\eqf
According to  \rfr{ring}, the precession induced by the asteroids' ring on the Venus'perihelion amounts to $+0.0007\pm 0.0001$ \asec\ by using $m_{\rm ring}=(5\pm 1)\times 10^{-10}$M$_{\odot}$ \citep{ringru}; the lowest value $+0.0006$ \asec\ is incompatible with \rfr{venu} at $10-\sigma$ level.
In the case of the Kuiper Belt Objects, \rfr{ring} yields a precession of the order of $+0.00006$ \asec\ with $m=0.052 m_{\oplus}$ \citep{IorKBO}.
Thus, we can rule out such modelled classical features of the Sun and the Solar System as explanations of $\Delta\dot\varpi_{\rm Venus}$.
General relativistic terms of order $\mathcal{O}(c^{-4})$ were not modelled by Pitjeva; however, the first correction of order $\mathcal{O}(c^{-4})$ to the perihelion precession \citep{Dam} can be safely neglected because for Venus it is
\eqi\dot\varpi_{c^4}\propto \rp{n (GM)^2}{c^4 a^2(1-e^2)^2}\approx 10^{-7}\ \ \ ''\ {\rm cy}^{-1}.\eqf

Concerning possible exotic explanations, i.e. due to some modifications of the currently known Newton-Einstein laws of gravity, it may have some interest to check some of the recently proposed extra-forces \citep{Sta08} which would be able to phenomenologically accommodate  the Pioneer anomaly \citep{Pio}. All of such hypothetical new forces have not been modelled by Pitjeva, so that if they existed in Nature they would affect $\Delta\dot\varpi_{\rm Venus}$.
A central acceleration quadratic in the radial component $v_r$ of the velocity of a test particle\footnote{The quoted numerical value of $\mathcal{H}$ allows to reproduce the Pioneer anomaly.} \citep{Jaek,Sta08}
\eqi A = -v_r^2{\mathcal{H}},\  {\mathcal{H}} = 6.07\times 10^{-18}\ {\rm m}^{-1}\eqf
would induce a retrograde perihelion precession according to \citep{IorPio08}
\eqi\dot\varpi = \rp{  {\mathcal{H}} na\sqrt{1-e^2}   }{e^2}\left(-2 + e^2 + 2\sqrt{1-e^2}\right)< 0.\lb{precpio}\eqf
However, \rfr{precpio} predicts a precession of $-0.0016$ $''$\ {\rm cy}$^{-1}$ for Venus, which is ruled out by \rfr{venu} at $12-\sigma$ level.
Another possible candidate considered in \citep{Sta08} is an acceleration linear in the radial velocity
\eqi A =-|v_r|{\mathcal{K}},\ \mathcal{K} = 7.3\times 10^{-14}\ {\rm s}^{-1}\eqf
which yields a retrograde perihelion precession  \citep{IorPio08}
\eqi\dot\varpi = -\rp{\mathcal{K}\sqrt{1-e^2}}{\pi}\left[\rp{2e - (1-e^2)\ln\left(\rp{1+e}{1-e}\right)}{e^2}\right]< 0.\lb{odot}\eqf
The prediction of \rfr{odot} for Venus is $-0.1$ $''$\ {\rm cy}$^{-1}$, clearly incompatible with \rfr{venu}.
Should one consider a central uniform acceleration with the magnitude of the Pioneer anomalous one, i.e. $A= -8.74\times 10^{-10}$ m s$^{-2}$,
the exotic precession induced by it \citep{IorPio,San06} on the perihelion of of Venus would be
\eqi\dot\varpi_{\rm Ven} = A\sqrt{\rp{a(1-e^2)}{GM}} = -16\ ''\ {\rm cy}^{-1}.\eqf
Another non-conventional effect which may be considered is the precession predicted by \citet{Lue} in the framework of the DGP multidimensional braneworld model by Dvali, Gabadadze and Porrati \citep{DGP} proposed to explain the cosmic acceleration without invoking dark energy. It is \eqi\dot\varpi_{\rm LS}=\mp\rp{3c}{8r_0}+ \mathcal{O}(e^2) \approx \mp 0.0005\ ''\ {\rm cy}^{-1}, \lb{digip}\eqf
where the plus sign is related to the self-accelerated branch, while the minus sign is for the standard, Friedmann-Lema\^{\i}tre-Robertson-Walker (FLRW) branch; $r_0\approx 5$ Gpc is a threshold characteristic of the DGP model after which gravity would experience neat deviations from the Newtonian-Einsteinian behavior. As can be noted, the self-accelerated branch is ruled out at $9-\sigma$ level by \rfr{venu}, while the FLRW case is still compatible with \rfr{venu} ($1-\sigma$ discrepancy).  By the way, apart from the fact that there are theoretical concerns with the DGP model  (see, e.g., \citep{Koy} and references therein), the existence of both the Lue-Starkman FLRW precession and the Lense-Thirring one, implying a total unmodelled effect of $-0.0008$ \asec,  would be ruled out by \rfr{venu} at $4-\sigma$ level.
As a consequence, we can conclude not only that the examined exotic modifications of the standard laws of gravity, not modelled by Pitjeva, are not responsible for the estimated $\Delta\dot\varpi_{\rm Venus}$, but also that their existence in the inner regions of the Solar System is falsified by the observations.
Moreover, given the magnitudes of the hypothetical effects with the negative sign, it is not possible that reciprocal cancelations with the positive classical mismodelled precessions can explain \rfr{venu}. Indeed, the sum of the latter ones is $+0.0004$ \asec; the sum of, e.g., \rfr{precpio} and \rfr{digip} (FLRW) is $-0.0021$ \asec, while the sum of \rfr{precpio} and \rfr{digip} (self-accelerated branch) is $-0.0011$ \asec.

Thus, we conclude that the most likely explanation for \rfr{venu} is just the general relativistic Lense-Thirring effect. However, caution is in order in assessing the realistic uncertainty in such a test because, as already stated, the released error of 0.0001 \asec\ is the formal, statistical one; the realistic uncertainty might be larger. By the way, we can at least firmly conclude that now also in the case of Venus the general relativistic predictions for the Lense-Thirring effect on $\dot\varpi$ are compatible with the observational determinations for the unmodelled perihelion precessions, contrary to the case of \citep{IorPS}. Moreover, future modelling of planetary motions should take into account the relativistic effects of the rotation of the Sun as well. The steady improvement in the planetary ephemerides, which should hopefully  benefit of the radiometric data from Messenger and Venus-Express as well, should allow for more accurate and stringent test in the near-mid future. Of great significance would be if also other teams of astronomers would estimate their own corrections to the canonical perihelion (and also node) precessions in order to enhance the statistical significance and robustness of this important direct test of general relativity.

\section*{Acknowledgments}
I thank E.V. Pitjeva for useful and important communications.


\end{document}